\documentclass{article}

\usepackage{microtype}
\usepackage{graphicx}
\usepackage{subfigure}
\usepackage{booktabs} 

\usepackage{amsmath}
\usepackage{url}
\usepackage{pifont}
\usepackage{amssymb}
\usepackage{enumitem}
\usepackage{seqsplit}
\usepackage{xcolor}
\usepackage{multirow}
\usepackage{makecell}

\usepackage{hyperref}
\newcommand{\sys}{\textsc{BOute}\xspace}

\usepackage[accepted]{mlsys2025}
\usepackage[nobiblatex]{xurl}
\usepackage{xspace}
\usepackage{ulem}

\begin{document}

\twocolumn[
\mlsystitle{\sys: Cost-Efficient LLM Serving with Heterogeneous LLMs and GPUs via Multi-Objective Bayesian Optimization}




\begin{mlsysauthorlist}
\mlsysauthor{Youhe Jiang}{to}
\mlsysauthor{Fangcheng Fu}{goo}
\mlsysauthor{Eiko Yoneki}{to}
\end{mlsysauthorlist}

\mlsysaffiliation{to}{Department of Computer Science, University of Cambridge, Cambridgeshire, UK}
\mlsysaffiliation{goo}{School of Artificial Intelligence, Shanghai Jiao Tong University, Shanghai, China}

\mlsyscorrespondingauthor{Fangcheng Fu}{ccchengff@sjtu.edu.cn}
\mlsyscorrespondingauthor{Eiko Yoneki}{eiko.yoneki@cl.cam.ac.uk}

\mlsyskeywords{Machine Learning, MLSys}

\vskip 0.3in

\begin{abstract}
The rapid growth of large language model (LLM) deployments has made cost-efficient serving systems essential. Recent efforts to enhance system cost-efficiency adopt two main perspectives: (\textbf{\underline{i}}) An \textit{algorithmic} perspective that exploits heterogeneous model capabilities to route simpler queries to lower-cost models and complex queries to higher-cost models (i.e., heterogeneous model routing); and (\textbf{\underline{ii}}) a \textit{systems} perspective that utilizes heterogeneous GPU resources as cost-effective alternatives to homogeneous high-end GPUs (i.e., heterogeneous model deployment). However, algorithm-system co-design for cost-efficient LLM serving necessitates sophisticated management: (\textbf{\underline{i}}) Determining optimal routing strategies under latency and quality requirements, (\textbf{\underline{ii}}) configuring model deployment across heterogeneous GPUs with appropriate resource allocation and parallelism strategies, and (\textbf{\underline{iii}}) co-optimizing routing and deployment decisions to maximize overall system performance. To address these challenges, we present \sys, a \textit{quality-aware scheduling system} that jointly exploits heterogeneous model and GPU capabilities for cost-efficient LLM serving. \sys employs a \textit{multi-objective Bayesian optimization (MOBO) framework} to co-optimize the routing strategy and model deployment, thereby maximizing the cost-efficiency of the serving system while guaranteeing response quality. 
Evaluation results demonstrate that \sys outperforms state-of-the-art LLM serving systems by up to 157\% and 59\% on average under \textit{identical} cost budgets and quality requirements, or reducing serving costs by 15\%-61\% (38\% on average) while maintaining the \textit{same} performance targets, validating its effectiveness in achieving cost-efficient LLM serving.
\end{abstract}
]



\printAffiliationsAndNotice{} 

\section{Introduction}
\label{sec:intro}
Large language models (LLMs) such as Llama-3~\cite{dubey2024llama}, DeepSeek-R1~\cite{guo2025deepseek}, Claude~\cite{anthropic2025claude}, Gemini~\cite{comanici2025gemini}, and GPT-5~\cite{openai2025gpt5} have demonstrated outstanding performance across diverse real-world applications, including chatbots, healthcare, and education, profoundly influencing human lives~\citep{jeon2023large,peng2023study,copilot}. However, LLM serving incurs substantial costs~\citep{jiang2025demystifying,miao2024spotserve}, as it demands significant computational resources (e.g., GPUs) to satisfy latency and quality requirements. Consequently, how to achieve cost-efficient LLM serving is a timely and essential topic for service providers.

Typically, when a user query arrives, service providers must make two essential decisions during the serving process: (\textbf{\underline{i}}) \textit{which model} should process the query (depending on query complexity and model capability) and (\textbf{\underline{ii}}) \textit{how to deploy} each model across available GPU resources (through resource allocation and parallelism strategies). The ultimate objective of jointly optimizing these two decisions is to achieve cost-efficient LLM serving while maintaining stringent latency and quality requirements. In this work, we explore heterogeneous model routing~\cite{ong2024routellm,ding2024hybrid} and heterogeneous model deployment~\cite{jiang2025thunderserve,mei2025helix} to leverage the diverse capabilities of heterogeneous models and GPU resources, thereby enhancing cost-efficiency in LLM serving.

Heterogeneous model routing is a serving paradigm that dynamically assigns incoming queries to the most appropriate model based on predicted response quality requirements. In this approach, a router evaluates each query,
and makes a routing decision in a single forward pass~\cite{ong2024routellm,ding2024hybrid}. Queries deemed easy or low-stakes are directed to smaller, more efficient models, while those identified as difficult or high-stakes are routed to larger, more capable models. By aligning query requirements with model capabilities while satisfying latency and quality requirements, this paradigm enables cost-efficient serving without compromising response quality. 
Several recent studies~\cite{hu2024routerbench,chuang2024learning,varangot2025doing,panda2025adaptive} have focused on cost-efficient LLM serving using heterogeneous model routing.

Recent efforts~\cite{jiang2025demystifying,griggs2024m,mao2025skyserve,mei2025helix} have also proposed leveraging heterogeneous GPU resources for cost-efficient LLM serving (i.e., heterogeneous model deployment). Unlike traditional homogeneous LLM deployments that rely exclusively on high-end GPUs (e.g., H100), this paradigm utilizes a mixture of GPU types, ranging from premium GPUs to more economical alternatives (e.g., RTX 5090), to create a cost-effective serving infrastructure. 

As the heterogeneity exists in both models and hardwares, we observe that \textit{heterogeneous model deployment naturally complements heterogeneous model routing}. In particular, in a routing system, since different models exhibit varying resource demands (e.g., compute requirements, memory footprints, and bandwidth utilization), it is intuitive to leverage different types of GPUs tailored to each model's characteristics~\cite{jiang2025demystifying,griggs2024m}. We present a detailed workload characterization in \S\ref{sec:workload}.

Algorithm-system co-design is necessary for heterogeneous model routing (algorithm) and deployment (system). This is due to the bidirectional dependence between these two components:
\textbf{\underline{First}}, routing decisions shape the system load distribution (e.g., request arrival rates) across heterogeneous models, directly influencing optimal model deployment strategies. \textbf{\underline{Conversely}}, deployment configurations determine the system latency achievable by each model, which in turn impact routing decisions. Therefore, achieving an efficient, quality-guaranteed serving system necessitates the joint optimization of routing strategy and model deployment. However, realizing such co-optimization in practice proves much harder to implement than to propose:
\begin{itemize}[leftmargin=*]
\vspace{-0.75em}
    \item \textbf{Effective model routing.} Determining an optimal routing strategy requires balancing two competing objectives: system response latency and quality. This decision is further complicated when leveraging heterogeneous resources for serving, where different LLMs on different GPU types exhibit varying system latency, making routing decisions inherently dependent on model deployment.
\vspace{-0.5em}
    \item \textbf{Efficient model deployment.} Different GPU types have distinct characteristics (e.g., compute capability, memory bandwidth, memory capacity), and different LLMs favor different allocations and parallelisms (e.g., data, tensor, pipeline parallelism). This makes model deployment optimization on heterogeneous GPU clusters a complex problem. Co-designing with model routing amplifies this complexity, as routing decisions determine cross-model system loads that influence optimal model deployment.
\vspace{-0.5em}
    \item \textbf{Algorithm-system co-design.} The mutual dependencies between routing and deployment create a challenging circular optimization problem. On the one hand, determining the optimal routing strategy requires knowing the latency and quality characteristics of a given model deployment. On the other hand, selecting the optimal model deployment requires knowing the system load distribution shaped by the routing strategy. This interdependence  implies that optimizing routing and deployment in isolation leads to suboptimal system performance.
\vspace{-0.75em}
\end{itemize}

In order to overcome these challenges, we propose \sys, a quality-aware scheduling system that jointly exploits heterogeneous model and GPU capabilities for cost-efficient LLM serving. Our contributions are summarized as follow:

\textbf{\underline{Contribution 1:}} We systematically characterize the system workload by exploring how co-designing routing and deployment impacts system performance, and demonstrate that heterogeneous model deployment naturally complements heterogeneous model routing: matching the resource demands of different LLMs to appropriate GPU types can substantially improve system efficiency.

\textbf{\underline{Contribution 2:}} We formulate the scheduling problem of routing and deployment across heterogeneous models and resources as a constraint optimization problem. To solve this problem efficiently, we propose a multi-objective Bayesian optimization (MOBO) framework to co-optimize routing strategy and model deployment. Specifically, given latency and quality objectives, the MOBO framework identifies Pareto-optimal solutions on the latency-quality frontier, enabling service providers to select deployments that best satisfy their performance requirements.

\textbf{\underline{Contribution 3:}} We implement \sys and conduct experiments across diverse LLM tasks and quality requirements. Compared to state-of-the-art LLM serving systems, \sys achieves up to 2.6$\times$ reduction in system response latency with 1.6$\times$ average improvement, or boosts system throughput by up to 1.9$\times$ (with 1.6$\times$ on average) given the \textit{same} serving budget and quality requirement. 

\section{Background and Related Works}
\label{sec:preliminaries}


\textbf{Heterogeneous model routing.} Model routing systems deploy a router model to dynamically direct incoming queries to different LLMs based on query characteristics, with the primary objective of optimizing system metrics such as cost, latency, or throughput while maintaining output quality. Several recent works~\cite{chenfrugalgpt,hu2024routerbench} have explored this paradigm: RouteLLM~\cite{ong2024routellm} develops cost-optimal routing strategies using preference data and calibrated confidence scores to balance performance and cost; HybridLLM~\cite{ding2024hybrid} proposes an adaptive LLM routing system that switches between large and small models based on query complexity. 
These works primarily focus on router design and routing strategies for query-level model selection, with limited attention to the routing impact on system performance. In contrast, our work, from a service provider perspective, addresses deployment-level optimization to further enhance system performance.

\textbf{Heterogeneous model deployment.} Heterogeneous model deployment refers to LLM serving systems~\cite{jiang2024hexgen,miao2024spotserve,mao2025skyserve} that utilize GPUs with varying computational capabilities, memory capacities, or availability characteristics to optimize resource utilization and cost efficiency. Among them, 
ThunderServe~\cite{jiang2025thunderserve} accommodates the different computational characteristics of prefill and decoding phases by deploying them on heterogeneous GPUs matched to their respective workload requirements; 
Helix~\cite{mei2025helix} formulates heterogeneous clusters as a flow network and utilizes a max-flow algorithm to maximize system throughput. 
Our work shares a similar motivation: utilizing heterogeneous resources for cost-efficient LLM serving, and we find it naturally complements heterogeneous model routing. A more detailed related work is demonstrated in~\autoref{appendix:relatedwork}.

\section{Workload Characterization}
\label{sec:workload}
This characterization focuses on understanding how heterogeneous model routing and deployment jointly impact system performance and cost efficiency. 
We introduce different approaches to improve the system performance step-by-step.

\textbf{Experiment setup.} We use P95 latency (the 95th percentile of response latencies) as our primary performance metric and evaluate quality (i.e., accuracy) on a sample of the GSM8K dataset. We compare two deployment configurations under similar budget constraints: a homogeneous setup with 12 H100 GPUs (\$32.28/h) and a heterogeneous setup with 6 RTX 5090 GPUs and 10 H100 GPUs (\$32.24/h).


\textbf{Baseline system configuration.} We begin by establishing a baseline scenario where a service provider must satisfy a minimum quality requirement of \textbf{90} (on a scale of 0-100). Consider a system deploying Llama3.1-70B on 12 H100 GPUs, which achieves a P95 latency of \textbf{25.6s} and a quality score of \textbf{95.1}. Alternatively, deploying only Llama3.1-8B on the same 12 H100 GPUs yields a P95 latency of \textbf{7.4s} but a quality score of \textbf{84.5}, failing to meet the required threshold. This illustrates the fundamental trade-off between latency and quality when serving individual models.

\textbf{\underline{Approach 1}: Model routing with uniform resource allocation.} Introducing model routing enables the system to satisfy quality requirements while reducing latency. With an appropriate routing strategy, approximately 40\% of queries are directed to the small model while 60\% are routed to the large model, achieving the target quality of \textbf{90}. And with uniform resource allocation (6 GPUs per model), the system achieves a P95 latency of \textbf{28.2s} (\textbf{10\%}$\uparrow$ compared to single Llama3.1-70B model deployment). 

\begin{figure}
    \centering
    \includegraphics[width=\linewidth]{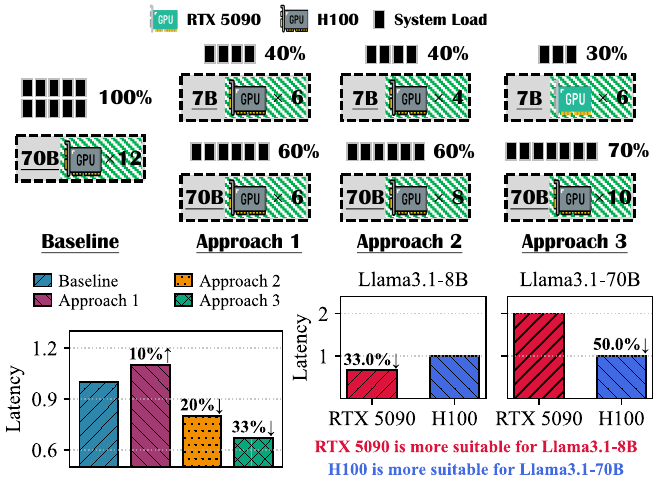}
    \caption{\textbf{\underline{Upper}}: Illustration of system load assignment and resource allocation for approaches 1-3. \textbf{\underline{Lower Left}}: Performance comparison among the baseline and 3 approaches. \textbf{\underline{Lower Right}}: Model performance across GPU types. The latency is normalized by each model's P95 latency on H100 GPUs.}
    \label{fig:charac_sys}
    \label{fig:charac}
\end{figure}

\textit{\underline{Insight:}} Despite routing reducing the load on the large model, inappropriate resource allocation creates a performance bottleneck. The large model, handling 60\% of queries with only half the available resources, becomes the limiting factor for overall system latency.

\textbf{\underline{Approach 2}: Optimizing homogeneous resource allocation.} The system performance can be substantially improved by adjusting resource allocation. Allocating 4 GPUs to the small model and 8 GPUs to the large model—based on their assigned system loads and model characteristics—reduces the P95 latency to \textbf{20.5s} (\textbf{20\%}$\downarrow$ compared to single Llama3.1-70B model deployment).

\textit{\underline{Insight:}} Model routing combined with appropriate resource allocation significantly reduces system latency. By directing easier queries to the small model and balancing resources between the small and large models to prevent bottlenecks, the system avoids overloading either model.

\textbf{\underline{Approach 3}: Introducing heterogeneous GPUs.} Further latency reduction is achievable by deploying heterogeneous GPU types matched to model characteristics. In the heterogeneous setup, we route 30\% and 70\% of queries to the small and large models, respectively, deploying 6 RTX 5090 GPUs for the small model and 10 H100 GPUs for the large model. This configuration achieves a P95 latency of \textbf{17.1s} (\textbf{33\%}$\downarrow$ and \textbf{13\%}$\downarrow$ compared to single Llama3.1-70B and homogeneous model deployment), and a quality score of \textbf{91.2} (higher quality). 
\autoref{fig:charac} demonstrates the performance comparison among the baseline and approaches 1-3.

\textit{\underline{Insight:}} Integrating heterogeneous GPU resources alters the optimal routing strategy by changing the latency-quality trade-off curves of individual models.
The routing distribution shifts from 40/60 to 30/70 for two reasons: First, the small model achieves lower latency at reduced cost on RTX 5090 GPUs; second, this cost saving frees up budget to provision more H100 GPUs for the large model, enabling it to handle increased system loads.
Heterogeneous model deployment thus creates a dual benefit—cost-efficient serving of the small model while expanding large model capacity—that jointly minimizes overall system latency.


\textbf{Model performance across GPU types.} To demystify the performance characteristics of different models across GPU types, we conduct additional experiments comparing small and large model deployments on RTX 5090 and H100 GPUs under identical budget constraints. Specifically, we compare deployments using 24 RTX 5090 GPUs (\$21.36/h) and 8 H100 GPUs (\$21.52/h), representing equivalent costs.
Our benchmarking reveals that models exhibit distinct performance profiles on different GPU types, as shown in~\autoref{fig:charac}. The small model achieves approximately \textbf{1.5}$\times$ lower P95 latency on RTX 5090 GPUs compared to H100 GPUs, while the large model achieves \textbf{2}$\times$ lower P95 latency on H100 GPUs compared to RTX 5090 GPUs.

\textit{\underline{Insight:}} Different models favor different GPU types for cost-efficient serving, making heterogeneous GPU deployment essential rather than optional. This hardware heterogeneity naturally complements heterogeneous model routing: routing exploits model diversity to balance latency and quality, while heterogeneous deployment exploits GPU diversity to optimize the cost-performance of each model, creating a synergistic system where both dimensions of heterogeneity work together to maximize overall serving cost-efficiency.

Based on these insights, we manage to improve the cost efficiency via heterogeneous model routing and model deployment co-optimization. 
We first present the problem formulation in \S\ref{sec:formulation} and introduce our solution in \S\ref{sec:mobo framework}.

\section{Formulation of Co-optimization}
\label{sec:formulation}

\subsection{Decision Variables in Model Routing}
\label{subsec:routing}

\textbf{Our router.} We adopt a threshold-based routing mechanism consistent with prior works~\cite{ong2024routellm,ding2024hybrid}. Specifically, for each incoming query, the router computes a routing score that estimates the query's difficulty or the likelihood that a smaller model can produce a satisfactory response. This score is then compared against configurable thresholds $\boldsymbol{\tau}=\{\tau_1,\tau_2,\dots\}$ to make the routing decision. When multiple models are available in the system, queries are routed to model $i$ if their routing score falls within the range $[\tau_{i-1}, \tau_i)$. 

\textbf{Decision variables.} The routing thresholds $\boldsymbol{\tau}$ constitute the primary decision variables for the routing strategy. These thresholds control the distribution of queries across models, directly impacting the system's latency-quality trade-off.
Higher threshold values route a greater proportion of queries to larger, more capable models, thereby ensuring higher output quality at the cost of increased system latency. Conversely, lower thresholds direct more queries to smaller models, reducing system latency while accepting potential quality degradation.
Through systematic optimization of $\boldsymbol{\tau}$, the routing strategy can be adapted to meet specific system objectives and performance constraints.

\subsection{Decision Variables in Model Deployment}
\label{subsec:deployment}

\textbf{Deployment configuration.} Given a set of available heterogeneous GPUs $\mathbf{D} = \{d_{1},d_{2},\dots,d_{N}\}$, where each $d_{n} \ge 0$ represents the number of GPUs of the $n$-th type, the deployment problem determines how to allocate these GPUs to serve $M$ model types. The deployment configuration must specify three key aspects: (\textbf{\underline{i}}) Model-GPU assignment: which GPU type(s) should be used to deploy each model, (\textbf{\underline{ii}}) resource allocation: how many GPUs of each type should be allocated to each model, and (\textbf{\underline{iii}}) parallelism strategy: the degree of data, tensor, and pipeline parallelism used to distribute each model across its allocated GPUs.

\textbf{Our deployment strategy.} Let there be $M$ model types. We formulate the deployment problem as finding an optimal allocation matrix $\mathbf{A} \in \mathbb{R}^{M \times N}$, where $a_{m,n}$ is the number of GPUs of type $n$ allocated to model type $m$, along with parallelism strategies $\mathbf{P} = \{p_1, p_2, \dots, p_M\}$, where $p_m$ specifies the optimal parallelism strategy for model type $m$. The allocation must satisfy two types of constraints: (\textbf{\underline{i}}) Resource constraints: $\sum_{m=1}^{M} a_{m,n} \leq d_n, \forall n$, i.e., the total allocation does not exceed available resources, and (\textbf{\underline{ii}}) budget constraints: $\sum_{m=1}^{M} \sum_{n=1}^{N} a_{m,n} \cdot b_n \leq B_{\text{cap}}$, where $b_n$ is the cost per unit time for GPU type $n$ and $B_{\text{cap}}$ is the total budget, ensuring that the deployment cost remains within the specified budget limit. 

\textbf{Decision variables.} The allocation matrix $\mathbf{A}$ and parallelism strategies $\mathbf{P}$ constitute the primary decision variables for the model deployment. 
Given a routing strategy $\boldsymbol{\tau}$ that determines the system load distribution $\mathbf{W}=\{\lambda_1,\lambda_2,\dots,\lambda_M\}$ among the $M$ model types, where $\lambda_m$ represents the system load for model type $m$, we systematically optimize $\mathbf{A}$ and $\mathbf{P}$ to minimize system latency subject to resource and budget constraints. 

\subsection{Problem Formulation}
\label{subsec:problem_formulation}

Given a set of $M$ model types and heterogeneous GPUs $\mathbf{D} = \{d_{1},d_{2},\dots,d_{N}\}$ with costs $\mathbf{B} = \{b_1, b_2, \dots, b_N\}$, we formulate the joint optimization as a multi-objective optimization problem:
\begin{equation}
\small
\begin{aligned}
\min_{\boldsymbol{\tau}, \mathbf{A}, \mathbf{P}} \quad & \left(L(\boldsymbol{\tau}, \mathbf{A}, \mathbf{P}), \, -Q(\boldsymbol{\tau})\right) \\
\text{s.t.} & \sum_{m=1}^{M} a_{m,n} \leq d_n \, (\forall n), \sum_{m=1}^{M} \sum_{n=1}^{N} a_{m,n} \cdot b_n \leq B_{\text{cap}}
\end{aligned}
\end{equation}
where $L(\cdot)$ is the system response latency (e.g., P95 latency), $Q(\cdot)$ is the aggregated output quality across all queries (e.g., accuracy on GSM8K), and $B_{\text{cap}}$ is the budget constraint. The decision variables are routing thresholds $\boldsymbol{\tau}$, GPU allocation matrix $\mathbf{A}$, and parallelism strategies $\mathbf{P}$. The goal is to find Pareto-optimal solutions that balance the trade-off between minimizing latency and maximizing quality.

This problem is challenging due to the coupling between routing and deployment: routing strategy $\boldsymbol{\tau}$ determines system load distribution across models, which affects required model deployment $(\mathbf{A}, \mathbf{P})$ (i.e., resource allocation and parallelism strategy), while model deployment determines each model's latency characteristics, which influences the optimal routing strategy. Additionally, the problem complexity is further exacerbated by the NP-hardness of heterogeneous resource allocation~\cite{yuan2022decentralized,jiang2024hexgen}. Combined with these factors, the vast search space makes exhaustive search computationally infeasible. Therefore, we design a Multi-Objective Bayesian Optimization (MOBO) framework tailored for efficiently solving this problem.

\section{The MOBO framework}
\label{sec:mobo framework}
\begin{figure}
    \centering
    \includegraphics[width=\linewidth]{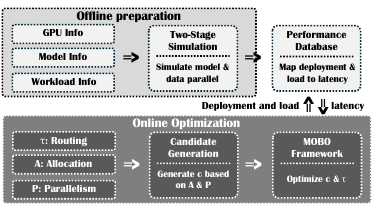}
    \caption{Illustration of the two phases in the MOBO framework.}
    \label{fig:workflow}
\end{figure}

As illustrated in~\autoref{fig:workflow}, our MOBO framework operates in two phases: (\textbf{\underline{i}}) Offline preparation phase: This phase profiles system performance and simulates different deployment configurations to generate performance data. This preparation is performed \textit{only} once for each specific serving scenario. (\textbf{\underline{ii}}) Online optimization phase: This phase leverages the offline performance data to efficiently search for optimal routing strategies and model deployments. 

\subsection{Offline Preparation Phase}
\label{subsec:mobo framework:p1}

The offline phase aims to map any $\langle$model deployment, system load$\rangle$ pair to an estimated system latency value.

\textbf{Inference task simulator.} An inference task simulator\footnote{We use the ETH EASL Scratchpad simulator~\cite{yao2023deltazip} to estimate system P95 latency. We show detailed simulator design (e.g., inputs, batching strategy, queuing mechanism, parallelism strategy modeling) and evaluation in~\autoref{appendix:simu}.} is employed to evaluate deployment performance. We profile heterogeneous GPU characteristics (e.g., effective compute throughput and achievable HBM I/O bandwidth) as inputs to the simulator, and iteratively simulate the performance of different model types under various model deployments $(\mathbf{A}, \mathbf{P})$ across a range of system loads $\lambda=\{2,4,6,8,\dots,40\}$ queries per second (QPS, spanning light to burst loads). The resulting performance data is then used in the subsequent online optimization phase.

\begin{figure}
    \centering
    \includegraphics[width=\linewidth]{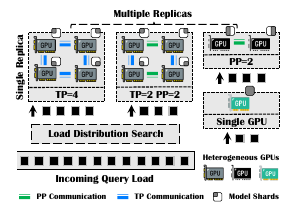}
    \caption{Illustration of single- and multi-replica deployments.}
    \label{fig:offline}
\end{figure}

\textbf{Two-stage simulation approach.} We adopt a two-stage simulation strategy based on parallelism types. 
\textbf{\underline{First}}, we simulate single-replica deployments using model parallelism, where a model is partitioned across GPUs. 
This constructs a performance database mapping $\langle$single-replica deployment, system load$\rangle$ to system latency. \textbf{\underline{Second}}, we simulate multi-replica deployments using data parallelism (DP)~\cite{li2023alpaserve}, where multiple replicas serve queries concurrently.
This maps $\langle$multi-replica deployment, system load$\rangle$ to system latency.
We demonstrate the illustration of single- and multi-replica deployments in~\autoref{fig:offline}.

\textbf{Simulation with model parallelism.} For each model type and available GPU type, we systematically enumerate feasible deployment configurations using model parallelism strategies. Model parallelism includes tensor parallelism (TP)~\cite{shoeybi2019megatron} and pipeline parallelism (PP)~\cite{huang2019gpipe}, which split a single model replica across multiple GPUs. Specifically, for GPU type $n$ with $d_n$ available units, we explore allocations from 1 to $d_n$ GPUs, evaluating all valid TP and PP combinations for each allocation size. 
The simulator generates the system latency for each specific model parallelism configuration.
To constrain the enumeration space, we apply two pruning rules: (\textbf{\underline{i}}) Configurations with insufficient GPU memory for the model are eliminated (e.g., 4 RTX 5090 GPUs for Llama3.1-70B), and (\textbf{\underline{ii}}) TP is restricted to utilize fast intra-machine GPU connections to avoid communication bottlenecks (e.g., TP $\leq$ 8 or 4, depending on hardware). This enumeration process is repeated across all model types, GPU types, and system loads, generating a performance database mapping model-parallel configurations to their corresponding system latencies. We demonstrate the enumeration complexity in~\autoref{appendix:offline simulation}.

\textbf{Simulation with data parallelism.} Beyond model parallelism, a model type may be deployed as multiple independent replicas using DP.
In this case, determining the optimal load distribution among these DP replicas remains non-trivial, as it directly impacts the final system latency.
To address this, we perform an efficient load distribution search: For each multi-replica configuration, we (\textbf{\underline{i}}) enumerate possible load distributions $(\lambda_{m,1}, \lambda_{m,2}, \dots)$ satisfying $\sum_i \lambda_{m,i} = \lambda_{m}$, where $\lambda_{m,i}$ denotes the system load assigned to replica $i$ of model type $m$, (\textbf{\underline{ii}}) retrieve the performance database for each replica's system latency under its assigned load, and (\textbf{\underline{iii}}) select the load distribution that minimizes the maximum system latency across different replicas, and record this minimum latency. Since each evaluation requires only database lookups, this search incurs minimal computational overhead. Through this approach, we can obtain the optimal load distribution and estimate system latency for arbitrary DP configurations.

\textbf{Summary.} The offline preparation phase constructs a comprehensive performance database that enables fast online optimization without real-time profiling or simulation overhead. Additionally, the offline preparation process typically completes within several dozen minutes and needs to be run only \textit{once} for each specific serving scenario.

\subsection{Online Optimization Phase}
\label{subsec:mobo framework:p2}
The online phase aims to utilize MOBO to find the Pareto-optimal routing strategies and model deployments.

\textbf{Deployment candidate generation.} 
We need to enumerate different deployments $(\mathbf{A}, \mathbf{P})$ for each model type. However, the number of possible candidates explodes with the number of GPU types, creating an extremely large search space for MOBO. Therefore, we generate deployment candidates for different model types as follows: 
\begin{itemize}[leftmargin=*]
\vspace{-0.75em}
\item \textbf{Homogeneous deployment candidates.} For each system load and GPU type, we enumerate feasible (DP, TP, PP) combinations and apply \textit{Pareto-skimming} in the (latency, cost) space to generate a set of Pareto-optimal configurations (typically 10-20 data points per GPU type with different latency-cost trade-offs). These configurations serve as homogeneous deployment candidates.
\vspace{-0.5em}
\item \textbf{Heterogeneous deployment candidates.} We generate heterogeneous candidates by combining homogeneous candidates from different GPU types. Specifically, we enumerate a grid of intermediate budgets $o \in (0,B_{\text{cap}}]$ (e.g., 10-20 points). For each budget $o$, we solve a \textit{multiple-choice knapsack problem} that selects at most one candidate from each GPU type to minimize system latency subject to the constraint that total cost $\leq o$. During knapsack solving, the system latency for each heterogeneous DP configuration is obtained through the load distribution search described in~\S\ref{subsec:mobo framework:p1}. The final set of configurations obtained across all budgets serves as heterogeneous deployment candidates.
\vspace{-0.75em}
\end{itemize}
Through this approach, we effectively cover the entire deployment space by pruning sub-optimal solutions via Pareto-skimming and knapsack optimization, reducing the candidate set to a manageable size without sacrificing solution quality. We denote the set of deployment candidates for model type $m$ as $\mathcal{C}_m$. The deployment choices are represented as $\mathbf{c} = \{c_1,c_2,\dots,c_M\}$, where $c_m \in \mathcal{C}_m$ for all $m \in \{1,\dots,M\}$.

\textbf{Overall MOBO workflow.}
The online optimization phase takes decision variables $\boldsymbol{\theta}=(\boldsymbol{\tau},\mathbf{c})$ as inputs and outputs the outcomes: response quality $Q(\boldsymbol{\theta})$\footnote{For different routing strategies $\boldsymbol{\tau}$, we use a sample of the incoming queries (i.e., a subsampled trace) to estimate their response quality $Q(\boldsymbol{\theta})$, following prior works~\cite{ong2024routellm,ding2024hybrid,shafran2025rerouting}.} and system latency $L(\boldsymbol{\theta})$\footnote{Aggregated P95 system latency across all model types.}. The MOBO framework primarily comprises two components: a surrogate model—a Gaussian Process (GP)—that provides fast statistical approximations of the objectives and constraints, and an acquisition function that balances exploration and exploitation over the decision space. The optimization workflow proceeds as follows: (\textbf{\underline{i}}) Surrogate model fitting: We fit GP surrogates for $Q(\boldsymbol{\theta})$ and $L(\boldsymbol{\theta})$ using additive kernels and structural information to capture the relationship between outcomes and decision variables. (\textbf{\underline{ii}}) Acquisition function optimization: We construct the constrained qNEHVI (q-Noisy Expected Hypervolume Improvement) acquisition function and optimize it over the decision space to balance exploration and exploitation, selecting the next candidate configuration $\boldsymbol{\theta}^*$ to evaluate. (\textbf{\underline{iii}}) Update and refinement: We evaluate the selected configuration $\boldsymbol{\theta}^*$, update the GP surrogates based on the observed performance, and repeat steps (\textbf{\underline{ii}}) and (\textbf{\underline{iii}}) until convergence.
Upon convergence, we obtain a \textit{Pareto-optimal set of solutions}, from which the optimal configuration is selected based on specific performance requirements (e.g., target system latency and response quality).

\textbf{Additive kernels.} To enable the surrogate model to capture both the separate effects and the interaction between routing strategies and deployment candidates, we employ a GP with additive kernels:
\[
\small
k(\boldsymbol{\theta},\boldsymbol{\theta}^*)
=
k_{\tau}(\boldsymbol{\tau},\boldsymbol{\tau}^*)
\;+\;
k_{c}(\mathbf{c},\mathbf{c}^*)
\;+\;
\phi\,k_{\times}\!\big((\boldsymbol{\tau},\mathbf{c}),(\boldsymbol{\tau}^*,\mathbf{c}^*)\big)
\]
where $k_{\tau}$ and $k_{c}$ are ARD Matérn kernels~\cite{snoek2012practical} that model the separable main effects of the routing strategies $\boldsymbol{\tau}$ and the deployment candidates $\mathbf{c}$, respectively. The interaction term $k_{\times}$ is a product kernel~\cite{duvenaud2013structure} that captures the interdependencies between $\boldsymbol{\tau}$ and $\mathbf{c}$. $\phi$ controls the weight of the interaction term and is learned during training (typically small).
This additive kernel structure allows the GP to disentangle main effects from interactions~\cite{kandasamy2015high}, yielding improved sample efficiency and interpretability~\cite{gardner2018gpytorch}.

\textbf{Structural information.} Appropriate structural information is essential for solving complex MOBO problems, as it guides the surrogate model and acquisition function toward feasible and promising regions. We incorporate several types of structural information based on algorithmic properties, system constraints, and empirical observations (S1-5):

\textit{\underline{S1.} Algorithmic reparameterization: Load-fraction encoding.} Rather than directly optimizing routing thresholds $\boldsymbol{\tau}$, we reparameterize them as load fractions—the proportion of system load assigned to each model type. This reparameterization provides several advantages: Small changes in raw thresholds can cause large, discontinuous jumps in load allocation due to non-uniform score distributions, whereas load fractions map linearly to system load and latency, stabilizing the optimization procedure and enabling natural enforcement of linear constraints (e.g., $\sum_{m=1}^M f_m = 1$, where $f_m$ is the load fraction assigned to model type $m$).

\textit{\underline{S2.} System constraint: Budget enforcement.} We enforce budget feasibility through hard constraints on total deployment cost. For any deployment decision $\mathbf{c}$, the accumulated cost across all model types must satisfy $\sum_{m=1}^{M} \sum_{n=1}^{N} a_{m,n} \cdot b_n \leq B_{\text{cap}}$, where $b_n$ is the cost per GPU of type $n$ and $B_{\text{cap}}$ is the budget limit. Configurations violating this constraint are excluded from the search space.

\textit{\underline{S3.} System constraint: GPU availability.} Similarly, we enforce hard constraints on GPU resource availability. For any deployment decision, the accumulated GPU allocation of each type must satisfy $\sum_{m=1}^{M} a_{m,n} \leq d_n$ for all GPU types $n$, where $d_n$ is the available count of GPU type $n$. Violations are excluded from the search space.

\textit{\underline{S4.} Output transformation: Scale normalization.} Due to the large magnitude differences between latency and quality objectives, we apply objective-specific transformations: system latency $L(\boldsymbol{\theta})$ is transformed to $\log(L(\boldsymbol{\theta}))$ to stabilize large variations, while response quality $Q(\boldsymbol{\theta})$ is transformed to $\log\left(\frac{Q(\boldsymbol{\theta})}{1-Q(\boldsymbol{\theta})}\right)$ (logit transformation) to linearize small variations near boundary values. These transformations improve GP conditioning and optimization stability.

\textit{\underline{S5.} Heterogeneous GPU awareness: Model-GPU preference encoding.} As demonstrated in~\S\ref{sec:workload}, different model types exhibit distinct performance characteristics on different GPU types. We encode this observation as a preference matrix $\mathbf{S}\in[0,1]^{M\times N}$, where $s_{m,n}$ represents the suitability of GPU type $n$ for model type $m$. This structural information is incorporated into the GP through a preference-weighted kernel: 
\[
\small
\tilde{k}_{c}(\mathbf{c}, \mathbf{c}^*) = w(\mathbf{c}) \cdot w(\mathbf{c}^*) \cdot k_{c}(\mathbf{c}, \mathbf{c}^*)
\]
where $k_{c}$ is the base deployment kernel and $w(\mathbf{c}) = \exp\{\beta \sum_{m=1}^{M}\sum_{n=1}^{N} a_{m,n} \cdot s_{m,n}\}$ is a weight function. Here, $a_{m,n}$ denotes the number of GPUs of type $n$ allocated to model type $m$ in configuration $\mathbf{c}$, and $\beta$ is a learnable parameter controlling the strength of the preference bias. This exponential weighting increases covariance between configurations using preferred GPU types, thereby biasing the GP toward favorable allocations. This structural bias improves sample efficiency by directing early exploration toward promising configurations while remaining data-overridable when other constraints (e.g., budget, availability) favor non-preferred GPU types.

\textbf{Constrained qNEHVI.} We employ constrained qNEHVI~\cite{daulton2021parallel} to incorporate specific performance constraints, such as $L(\boldsymbol{\theta}) \leq L_{\max}$ for system latency and $Q(\boldsymbol{\theta}) \geq Q_{\min}$ for response quality, where $L_{\max}$ and $Q_{\min}$ denote maximum latency and minimum quality requirements, respectively. We reformulate these constraints as feasibility conditions:
{\small
\begin{align*}
g_1(\boldsymbol{\theta}) = L(\boldsymbol{\theta}) - L_{\max} \leq 0, \quad
g_2(\boldsymbol{\theta}) = Q_{\min} - Q(\boldsymbol{\theta}) \leq 0
\end{align*}
}%
We leverage the joint posterior distribution of the GP surrogates for $L(\boldsymbol{\theta})$ and $Q(\boldsymbol{\theta})$ to evaluate constraint satisfaction. The constrained qNEHVI acquisition function computes expected hypervolume improvement by restricting Monte Carlo integration to feasible samples~\cite{balandat2020botorch}:
\[
\small
\text{Acq}(\boldsymbol{\theta}) = \mathbb{E}\left[\text{HVI}(\boldsymbol{\theta}) \cdot \mathbb{I}\left[\bigcap_{i=1}^{2} \{g_i(\boldsymbol{\theta}) \leq 0\}\right]\right],
\]
where the expectation is taken over samples from the joint posterior of $(L(\boldsymbol{\theta}), Q(\boldsymbol{\theta}))$, and $\mathbb{I}[\cdot]$ is the indicator function for feasibility. This formulation preserves the correlations between objectives and constraints during sampling, ensuring that the optimizer naturally balances Pareto frontier exploration with constraint satisfaction~\cite{gardner2014bayesian}.

\textbf{Summary.} The online phase employs a MOBO framework that efficiently searches the joint space of routing strategies and model deployments. By incorporating additive kernels, structural information (e.g., load-fraction encoding, model-GPU preference encoding), and enforcing hard constraints via constrained qNEHVI, the framework converges to Pareto-optimal solutions that balance system latency and response quality under given constraints and requirements.

\section{System Implementation}
\label{sec:impl}


\begin{figure*}
\centering
\includegraphics[width=\linewidth]{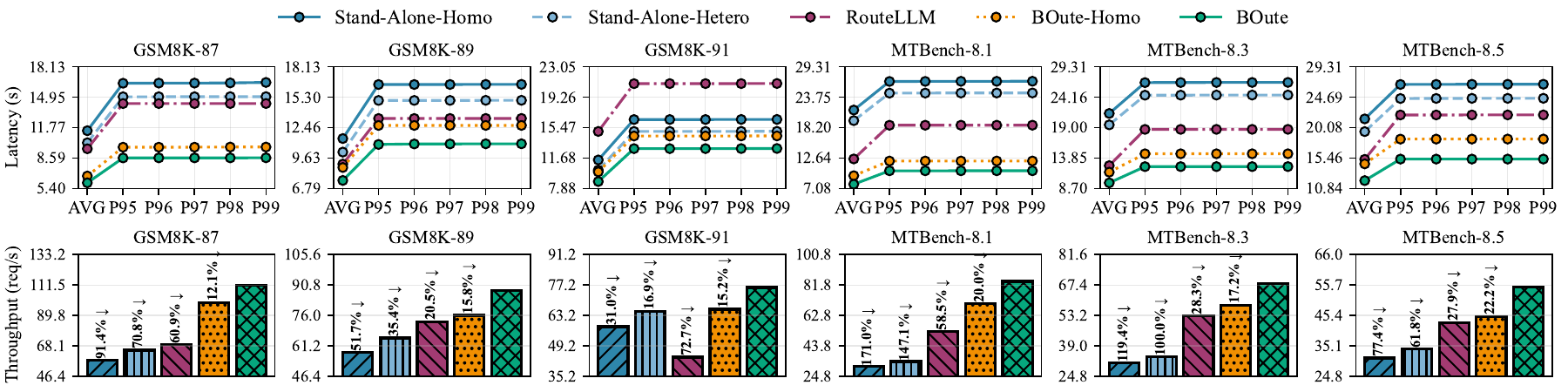}
    \caption{Experimental results of \sys compared with different baselines on GSM8K and MTBench workloads under different quality requirements. We select the minimum P95 latency routing strategy and model deployment from the Pareto-optimal solution set. GSM8K-87 represents a quality requirement of 87 (in terms of aggregated accuracy), MTBench-8.1 represents a quality requirement of 8.1 (in terms of aggregated score), and so on.}
    \label{fig:gsm8k_and_mtbench}
\end{figure*}

\textbf{Overall routine.} The overall routine of \sys operates as follows: To launch the serving process, the MOBO framework generates the optimal routing strategy and model deployment, which are then used to configure the router~\cite{ong2024routellm} and instantiate heterogeneous models across heterogeneous GPU resources. During serving, incoming queries are processed by a coordinator~\cite{yaoopen} (detailed in~\autoref{appendix:sys impl}), which connects to all models and dispatches queries to appropriate ones based on the guidance of scheduling results; generated responses are subsequently collected and returned to the coordinator.


\textbf{MOBO optimization kernels.} 
We implement several optimization kernels using GPyTorch~\cite{gardner2018gpytorch} for an efficient MOBO framework: (\textbf{\underline{i}}) An ARD \emph{Matérn–5/2} kernel (\texttt{MaternKernel} with \texttt{nu=2.5}) over routing features ($k_\tau$); (\textbf{\underline{ii}}) an ARD \emph{Matérn–3/2} kernel (\texttt{MaternKernel} with \texttt{nu=1.5}) over deployment features ($k_c$); (\textbf{\underline{iii}}) a low-weight interaction term $k_{\times}$ constructed via \texttt{ProductKernel(k\_$\tau$, k\_c)}; and (\textbf{\underline{iv}}) a \emph{custom} \texttt{PreferenceWeightedKernel} that wraps $k_c$ to inject structural preferences for model-to-GPU assignments. Each kernel is wrapped with a \texttt{ScaleKernel} to learn kernel-specific output scales, while ARD lengthscales are learned jointly by maximizing the Gaussian process marginal likelihood. The composite kernel ($k_\tau + \tilde{k}_c + k_{\times}$) is integrated into BoTorch~\cite{balandat2020botorch} \texttt{SingleTaskGP} models for system latency and response quality, which are combined via \texttt{ModelListGP}. Outcome transforms are applied as follows: \texttt{Log}$+$\texttt{Standardize} for system latency and \texttt{Logit}$+$\texttt{Standardize} for response quality.

\section{Evaluation}
\label{sec:exp}

\subsection{Experimental Setup}
\label{subsec:setup}

\textbf{Environments.} Our experiments are conducted on 4 types of GPU servers, each with 8 GPUs: NVIDIA H100-80G (\$2.64-3.07/h), RTX PRO 6000-96G (\$1.84-2.09/h), RTX 5090-32G (\$0.89/h), and RTX 4090-24G (\$0.59/h) GPUs. Within H100 servers, the GPUs are connected via NVLink with a bandwidth of 300GB/s; within RTX 5090 and 4090 servers, the GPUs are connected via PCIe with a bandwidth of 60GB/s. We compare homogeneous and heterogeneous model deployments under the same price budget of \$30/h.

\textbf{Models, router, and traces.} We employ Llama3.1-8B (smaller) and Llama3.1-70B (larger) as two model types in our system, which are representative and popular open-source transformer models~\cite{dubey2024llama}. For model routing, we adopt the router from~\cite{ong2024routellm}, which is trained on the Chatbot Arena dataset augmented with GPT-4-as-a-judge labeled data and golden-labeled MMLU validation data. The routing overhead remains negligible, accounting for less than 1\% of the total inference cost. And we follow prior work to generate workload traces based on real-world data~\cite{peng2025hexgen,zhong2024distserve}. Our testing traces are subsampled from GSM8K~\cite{cobbe2021training} and MTBench~\cite{zheng2023judging}.

\textbf{Evaluation metrics.} Following previous evaluation setups~\cite{miao2024spotserve}, we evaluate system performance based on
overall throughput and various percentile latencies (i.e., average, P95, $\dots$, P99, P100 latencies). In particular, the P95 latency denotes the maximum response time within which 95\% of all requests are completed.

\textbf{Baselines.} We compare \sys against four baseline systems:
(\textbf{\underline{i}}) \textbf{Stand-Alone-Homo}: vLLM~\cite{kwon2023efficient} serving the Llama3.1-70B model with homogeneous resources (H100 GPUs);
(\textbf{\underline{ii}}) \textbf{Stand-Alone-Hetero}: vLLM serving the Llama3.1-70B model with heterogeneous resources;
(\textbf{\underline{iii}}) \textbf{RouteLLM}: RouteLLM~\cite{ong2024routellm}, a state-of-the-art open-source framework for LLM routing, routing queries between Llama3.1-8B and Llama3.1-70B models based on query complexity, where each model is served by vLLM with uniform resource allocation across model types;
(\textbf{\underline{iv}}) \textbf{\sys-Homo}: \sys serving Llama3.1-8B and Llama3.1-70B models with routing and deployment optimization on homogeneous resources.
We compare the system performance of \sys with these baselines under quality requirements of 86, 88, and 90 for GSM8K and 8.1, 8.3, and 8.5 for MTBench. The system load is set to 100 req/s in total. For fair comparison, we tune the model deployment for each baseline under price budgets and report the best values in all experiments.

\subsection{End-to-end Evaluation}
\label{subsec:e2e}


\textbf{End-to-end system performance on GSM8K workloads.} As demonstrated in Figure~\ref{fig:gsm8k_and_mtbench} (left), we evaluate \sys with heterogeneous model deployment against baseline methods to demonstrate its effectiveness in achieving cost-efficient LLM serving on the GSM8K workload:

(\textbf{\underline{i}}) Compared to standalone LLM serving using Llama3.1-70B with vLLM (Homo and Hetero), \sys achieves superior performance across all quality requirements, reducing P95 latency by up to 91\% (57\% on average) and improving throughput by up to 90\% (55\% on average). Standalone LLM serving fails to leverage heterogeneous model resources, naively routing queries of varying complexity to a single model type, which results in either poor relative quality or excessive P95 latency.
(\textbf{\underline{ii}}) Compared to RouteLLM, \sys reduces P95 latency by up to 66\% (50\% on average) and improves throughput by up to 71\% (52\% on average). RouteLLM with uniform resource allocation across model types fails to account for the distinct characteristics of different model types, as larger models typically require more computing resources even when assigned fewer workloads. For instance, on the GSM8K trace with a quality requirement of 91, RouteLLM achieves even lower throughput than standalone deployment, as 70\% of the queries are processed by the large model with uniform resource allocation. (as detailed in~\S\ref{sec:workload}).
(\textbf{\underline{iii}}) Compared to \sys-Homo, \sys reduces P95 latency by up to 15\% (14\% on average) and improves throughput by up to 16\% (14\% on average). Although \sys-Homo incorporates adaptive resource allocation based on system load, it fails to fully exploit the GPU preferences of different model types (as discussed in~\S\ref{sec:workload}), thereby limiting cost-efficiency.

In contrast to these approaches, \sys jointly optimizes both the routing strategy (determining which model serves which query) and the deployment strategy (determining which GPU serves which model), thereby achieving superior cost-efficiency in LLM serving.


\textbf{End-to-end system performance on MTBench workloads.} We conduct additional experiments to evaluate the performance of \sys on MTBench workloads with varying quality requirements. As shown in Figure~\ref{fig:gsm8k_and_mtbench} (right), \sys achieves up to 157\% (115\% on average) performance improvement compared to standalone LLM serving (Homo and Hetero), up to 80\% (42\% on average) performance improvement compared to RouteLLM, and up to 21\% (20\% on average) performance improvement compared to \sys-Homo. These results demonstrate the comprehensive cost-efficiency of \sys across diverse workloads.

\begin{table}[!t]
\centering
\caption{Routing strategies and resource allocations of \sys used in~\autoref{fig:gsm8k_and_mtbench}.
}
\resizebox{\linewidth}{!}{%
\begin{tabular}{l | c | c | c | c}
\hline
\textbf{Trace-Quality} & \textbf{Model Type} & \textbf{Load} & \textbf{Budget} & \textbf{Allocated Resources} \\
\hline
\multirow{2}{*}{GSM8K-87} & Llama3.1-8B & 73.45\% & 38.58\% & 8$\times$5090+5$\times$4090 \\
 & Llama3.1-70B & 26.55\% & 61.42\% & 6$\times$H100 \\
\hline
\multirow{2}{*}{GSM8K-89} & Llama3.1-8B & 43.95\% & 17.86\% & 6$\times$5090 \\
 & Llama3.1-70B & 56.05\% & 82.14\% & 8$\times$H100 \\
\hline
\multirow{2}{*}{GSM8K-91} & Llama3.1-8B & 29.94\% & 9.28\% & 3$\times$5090 \\
 & Llama3.1-70B & 70.06\% & 90.72\% & 6$\times$H100+4$\times$6000 \\
\hline
\multirow{2}{*}{MTBench-8.1} & Llama3.1-8B & 80.58\% & 46.34\% & 8$\times$5090+8$\times$4090+6000 \\
 & Llama3.1-70B & 19.42\% & 53.66\% & 4$\times$H100+2$\times$6000 \\
\hline
\multirow{2}{*}{MTBench-8.3} & Llama3.1-8B & 61.17\% & 24.44\% & 8$\times$5090 \\
 & Llama3.1-70B & 38.83\% & 75.56\% & 6$\times$H100+2$\times$6000 \\
\hline
\multirow{2}{*}{MTBench-8.5} & Llama3.1-8B & 41.75\% & 12.04\% & 4$\times$5090 \\
 & Llama3.1-70B & 58.25\% & 87.96\% & 6$\times$H100+4$\times$6000 \\
\hline
\end{tabular}%
}
\label{tab:schedulingbenchmark}
\end{table}

\subsection{Scheduling Evaluation}
\label{subsec:casestudy}

\textbf{Scheduling results.} \autoref{tab:schedulingbenchmark} presents the scheduling outcomes of our MOBO framework, yielding several notable insights into routing and resource allocation strategies:
(\textbf{\underline{i}}) Load distribution adapts to quality requirements. As quality requirements increase, the framework progressively shifts system load and computational budget toward the Llama3.1-70B model. For example, when evaluated on the GSM8K trace with a quality threshold of 87, the small and large models handle 73\% and 27\% of the system load, respectively. However, when the quality requirement is raised to 91, this distribution shifts to 30\% and 70\%, reflecting the increased reliance on the more capable large model to meet stringent quality constraints.
(\textbf{\underline{ii}}) Resource/Budget allocation reflects model-specific resource requirements. The framework allocates computational resources based not only on assigned system load but also on intrinsic model characteristics (e.g., compute and memory demands). For instance, when tested on MTBench with a quality requirement of 8.3, the small and large models are assigned 61\% and 39\% of the system load, respectively. However, only 24\% of the total budget is allocated to the small model, as it requires substantially fewer computational and memory resources to achieve comparable performance. To prevent the large model from becoming a system bottleneck, the framework allocates 76\% of the budget to it, ensuring balanced system performance across both model types. (\textbf{\underline{iii}}) Heterogeneous GPU allocation optimizes resource utilization. The framework strategically assigns heterogeneous GPU resources to different model types based on their computational requirements and GPU availability. For example, when tested on MTBench with a quality requirement of 8.1, all RTX 5090 and 4090 GPUs are allocated to the small model, supplemented by one RTX PRO 6000 GPU due to availability constraints. Conversely, the large model are deployed on four H100 GPUs and two RTX PRO 6000 GPUs, which offer superior compute and memory capabilities aligned with the model's higher resource demands.
These results verify that \sys is able to take into account the heterogeneity of models and hardware, quality requirements, as well as budget and GPU constraints, generating routing strategies and model deployments that maximize system performance.

\begin{table}[!t]
\centering
\caption{The scheduling (algorithm converge) time and scalability of \sys. \sys (w/o structural info) disables algorithmic reparameterization, output transformation, and heterogeneous GPU awareness. \sys (w/o offline prep) disables the performance database. The algorithm is considered converged when the solution remains stable for more than 20 consecutive iterations.}
\resizebox{\linewidth}{!}{%
\begin{tabular}{c | c | c | c}
\hline
\textbf{\makecell{GPU Types \\ / GPU Number}} & \textbf{\sys} & \textbf{\makecell{\sys \\ (w/o structural info)}} & \textbf{\makecell{\sys \\ (w/o offline prep)}} \\
\hline
3 / 24 & 23.5 s & 3.6 min & $\approx$ 12 min \\
\hline
4 / 32 & 32.5 s & 5.3 min & $\approx$ 20 min \\
\hline
5 / 40 & 48.6 s & 7.1 min & $>$ 30 min \\
\hline
6 / 48 & 1.2 min & 8.9 min & $>$ 30 min \\
\hline
\end{tabular}
}
\label{tab:schedulingtime}
\end{table}

\textbf{Scheduling time.} We demonstrate the scheduling time of \sys across different cluster scales. As shown in~\autoref{tab:schedulingtime}, \sys requires less than 30 seconds to converge to the Pareto-optimal solutions when scheduling on 3 GPU types with 24 total GPUs. As the number of GPU types and total GPUs increases, the scheduling time increases near-linearly, demonstrating its strong scalability across different cluster scales.
To further validate the effectiveness of injecting structural information, we evaluate \sys (w/o structural info) that disables the structural information described in~\S\ref{subsec:mobo framework:p2}. Our experiments show that without structural information, \sys requires approximately 8$\times$ more BO evaluations to converge. Consequently, the overall scheduling time increases from under a minute to 4-9 minutes for 3-6 GPU types (24-48 total GPUs).
Finally, we evaluate \sys (w/o offline prep) to validate the effectiveness of our offline preparation phase. Without offline preparation, the scheduling time increases dramatically to 12-30+ minutes for 3-6 GPU types (24-48 total GPUs). This substantial increase occurs because each Bayesian optimization evaluation must re-run the simulator to obtain performance metrics for different candidate deployments, making the optimization process significantly more expensive.

\subsection{Case Study}
\label{ablationstudy}

\begin{figure}
    \centering
    \includegraphics[width=\linewidth]{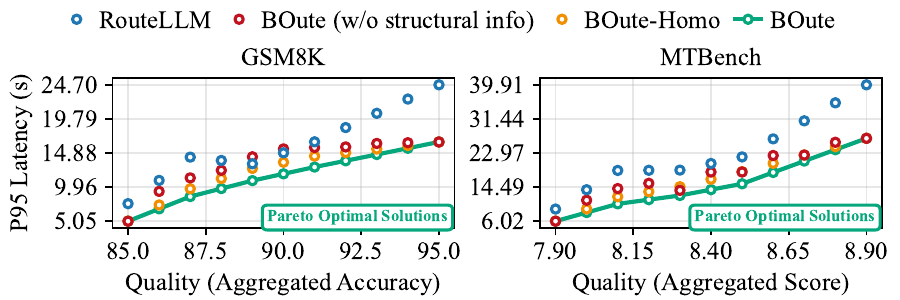}
    \caption{P95 latency results 
    across different quality requirements.}
    \label{fig:pareto-optimal solutions}
\end{figure}

\textbf{Pareto optimality of \sys.}
To further validate our approach, we evaluate \sys across varying quality requirements and compare its P95 latency against \sys-Homo, \sys (w/o structural info), and RouteLLM. As illustrated in~\autoref{fig:pareto-optimal solutions}, \sys consistently achieves superior P95 latency performance across different quality requirements.
Specifically, we observe the following: 
(\textbf{\underline{i}}) \sys exhibits Pareto dominance over \sys-Homo through its integration of heterogeneous resources and effective scheduling. Under equivalent quality requirements, \sys achieves lower latency; conversely, for a given latency constraint, \sys delivers higher quality. (\textbf{\underline{ii}}) \sys similarly demonstrates Pareto dominance over \sys (w/o structural info). Despite the additional scheduling overhead mentioned in~\S\ref{subsec:casestudy}, \sys without structural information frequently converges to local optima, resulting in inferior performance. In most cases, \sys (w/o structural info) is worse than not only \sys but also \sys-Homo. (\textbf{\underline{iii}}) RouteLLM with uniform resource allocation demonstrate the worst performance across different quality requirements.
These results collectively (\textbf{\underline{i}}) demonstrate the efficacy of integrating heterogeneous resources, (\textbf{\underline{ii}}) underscore the critical importance of incorporating structural information into the scheduling process, and (\textbf{\underline{iii}}) highlight the importance of resource allocation optimization in a routing system.

\begin{table}[!t]
\centering
\caption{
Costs required to meet quality and latency requirements.
(87, 8) represents quality and latency requirements of 87 and 8s.}
\resizebox{\linewidth}{!}{%
\begin{tabular}{l | c | c | c | c}
\hline
\textbf{Baseline} & \textbf{\makecell{GSM8K \\ (87, 8)}} & \textbf{\makecell{GSM8K \\ (91, 12)}} & \textbf{\makecell{MTBench \\ (8.1, 10)}} & \textbf{\makecell{MTBench \\ (8.5, 15)}} \\
\hline
Stand-Alone-Homo & \$61.83/h & \$41.21/h & \$79.93/h & \$53.29/h \\
\hline
Stand-Alone-Hetero & \$56.24/h & \$37.50/h & \$73.54/h & \$49.02/h \\
\hline
RouteLLM & \$53.55/h & \$52.38/h & \$55.90/h & \$44.00/h \\
\hline
\sys-Homo & \$38.41/h & \$36.07/h & \$36.33/h & \$38.31/h \\
\hline
\sys & \textbf{\$32.25/h} & \textbf{\$32.18/h} & \textbf{\$30.98/h} & \textbf{\$31.23/h} \\
\hline
\end{tabular}
}
\label{tab:performance}
\end{table}

\textbf{Cost-efficiency of \sys.} We evaluate the cost-efficiency of \sys under specified latency and quality requirements. Under equivalent quality constraints, systems typically require increased GPU resource allocation (higher budget) to achieve lower latency. \autoref{tab:performance} presents comparative results across multiple benchmark traces under varying quality and P95 latency requirements. \sys achieves the same latency and quality targets while reducing costs by an average of 39.7\% compared to Stand-Alone baselines and 38.0\% compared to RouteLLM. Moreover, \sys demonstrates a 15.0\% cost reduction relative to \sys-Homo, underscoring the advantages of heterogeneous model deployment. These results validate the effectiveness of \sys in reducing operational costs for real-time LLM serving systems.

\section{Conclusion}
This work focuses on cost-efficient LLM serving with heterogeneous models and GPUs. It proposes \sys, a serving system that employs a multi-objective Bayesian optimization (MOBO) framework to jointly optimize quality and latency through adaptive routing decisions for different queries and strategic model deployment across diverse models and hardware. Results show that, compared with baselines, \sys achieves up to 2.57$\times$ lower latency (1.58$\times$ on average) under identical cost budgets and quality requirements, or alternatively reduces costs by 15\%-61\% while maintaining the same performance targets. These results demonstrate \sys's cost-efficiency, and we believe it can contribute to the democratization of LLM serving.

\nocite{zhangsurvey,zhang2025turbodiffusion,zhang2025sla,zhang2024sageattention}
\bibliography{example_paper}
\bibliographystyle{mlsys2025}


\clearpage

\appendix

\section{Simulator Design and Validation}
\label{appendix:simu}

Our simulator employs a round-robin strategy for request dispatching among multiple parallel models, and a first-come first-served strategy for per-model request processing. The single-GPU processing time is based on profiled characteristics like compute TFLOPS and memory bandwidth. The simulator also considers the phase-specific characteristics of LLMs. The prefill phase is compute-bound, so its batched processing capacity is determined by the sum of the individual latencies. In contrast, the decoding phase is memory-bound, and its batched processing capability is defined by a single latency value. This distinction has been validated in several studies (e.g., DistServe~\citep{zhong2024distserve}, Splitwise~\citep{patel2024splitwise}).

\textbf{Simulator inputs.} In addition to GPU characteristics, the simulator requires a subsampled LLM workload trace. The workload trace is essential for accurate performance modeling, as it provides workload-specific characteristics such as input and output sequence length distributions that directly impact system performance.


\begin{table}[t!]
\centering
\caption{Simulator accuracy across parallelism configurations on Llama3-70B model under a workload with average input and output lengths of 1600 and 16. Errors are absolute percentage errors.}
\label{tab:sim-accuracy}
\resizebox{0.45\textwidth}{!}{%
\begin{tabular}{lccc}
\hline
\textbf{Config (DP,TP,PP)} & \textbf{Real (req/s)} & \textbf{Estimated (req/s)} & \textbf{Abs.\ \% Error} \\
\hline
(1, 4, 1) & 0.21 & 0.219 & 4.29\% \\
\hline
(1, 2, 2) & 0.26 & 0.280 & 7.69\% \\
\hline
(1, 1, 4) & 0.27 & 0.287 & 6.30\% \\
\hline
(2, 1, 2) & 0.33 & 0.347 & 5.15\% \\
\hline
(2, 2, 1) & 0.40 & 0.408 & 2.00\% \\
\hline
(2, 4, 1) & 0.41 & 0.437 & 6.59\% \\
\hline
(2, 2, 2) & 0.55 & 0.559 & 1.64\% \\
\hline
\end{tabular}%
}
\end{table}

\textbf{Batching strategy in our simulator.} The simulator's internal batching strategy is continuous batching, which iteratively batches request tokens to fully utilize the current resources. The GPU's memory limit constrains the maximum batch size for continuous batching.

\textbf{Queuing mechanism.} Our simulator maintains an individual queue for each model. Once there is free memory on the GPU (one request has finished), the model will fetch the next request in the queue for processing.

\textbf{Different parallelism.} Tensor and pipeline parallelism both split the computation workload of a single model across multiple devices. For pipeline parallelism, the simulator models communication overhead by profiling the relationship between estimated communication volume and observed latency. For tensor parallelism, the simulator assumes that each operator’s computation cost ideally scales down by a factor of $1/N$ when split across $N$ GPUs, and then adjusts this ideal cost using a speed-up coefficient $K(N)$ obtained from micro-benchmarks to account for communication and synchronization overhead. All profiling is performed offline before scheduling begins.

\textbf{Simulator evaluation.} We present the accuracy of our simulator with real-time experiments in~\autoref{tab:sim-accuracy}. The table presents examples of our throughput estimation for the Llama3-70B model under a workload with average input and output lengths of 1600 and 16, respectively. The notation (1,2,2) indicates a DP degree of 1, TP degree of 2, and PP degree of 2. Although the estimations are not perfectly accurate, they are sufficiently reliable (with estimation errors within 2\%–7\%) for selecting optimal configurations.

\section{Complexity Analysis of First-Stage Offline Simulation}
\label{appendix:offline simulation}

For each model type and available GPU type, we systematically enumerate feasible deployment configurations using model parallelism strategies. Model parallelism includes tensor parallelism (TP)~\cite{shoeybi2019megatron} and pipeline parallelism (PP)~\cite{huang2019gpipe}, which split a single model replica across multiple GPUs. Specifically, for GPU type $n \in \{1, \ldots, N\}$ with $d_n$ available units, we explore allocations from 1 to $d_n$ GPUs, evaluating all valid TP and PP combinations for each allocation size. 
For example, an 8-GPU allocation is simulated under configurations (TP=1, PP=8), (TP=2, PP=4), (TP=4, PP=2), and (TP=8, PP=1).
The enumeration complexity is $\Theta\left(|\lambda| M \sum_{n=1}^{N} d_n \log d_n\right)$,
where $|\lambda|$ represents the number of enumerated system load values, and the logarithmic factor accounts for valid TP and PP combinations per GPU allocation size.
After pruning, the enumeration complexity can be reduced to $\Theta\left(|\lambda| M \sum_{n=1}^{N} d_n\right)$.

\section{Extended System Implementation}
\label{appendix:sys impl}
\textbf{Overall routine.} The overall routine of \sys operates as follows: To launch the serving process, the MOBO framework generates the optimal routing strategy and model deployment, which are then used to configure the router and instantiate heterogeneous models across heterogeneous GPU resources. During serving, incoming queries are processed by a coordinator, which connects to all models and dispatches queries to appropriate ones based on the guidance of scheduling results; generated responses are subsequently collected and returned to the coordinator.

\textbf{Inference task coordinator.} To deploy \sys in a heterogeneous environment, we employ a task coordinator that incorporates a router responsible for generating routing scores for input queries~\cite{ong2024routellm,ding2024hybrid} and dispatches queries based on both these scores and the optimal routing strategy obtained from our scheduling algorithm. The task coordinator is primarily built upon an open-source implementation of decentralized computation coordination~\cite{yaoopen} that utilizes libP2P~\cite{libp2p} to establish peer-to-peer network connections between the coordinator and distributed models. 
When the task coordinator receives an input query, the routing process proceeds in three steps: \textbf{\underline{First}}, the router evaluates the query to determine the appropriate model type based on its routing score. \textbf{\underline{Second}}, the coordinator selects a specific replica of the chosen model type by consulting the local distribution search results described in~\S\ref{subsec:mobo framework:p1}, which balance workloads across replicas according to their heterogeneous deployment capabilities. \textbf{\underline{Third}}, the coordinator directs the query to the selected replica for execution and response generation.

\textbf{MOBO optimization kernels.} 
We implement several optimization kernels using GPyTorch~\cite{gardner2018gpytorch} for an efficient MOBO framework: (\textbf{\underline{i}}) an ARD \emph{Matérn–5/2} kernel (\texttt{MaternKernel} with \texttt{nu=2.5}) over routing features ($k_\tau$) to capture smooth, continuous effects of threshold variations; (\textbf{\underline{ii}}) an ARD \emph{Matérn–3/2} kernel (\texttt{MaternKernel} with \texttt{nu=1.5}) over deployment features ($k_c$) to accommodate rougher, step-like transitions when switching between deployment candidates; (\textbf{\underline{iii}}) a low-weight interaction term $k_{\times}$ constructed via \texttt{ProductKernel(k\_$\tau$, k\_c)} to capture cross-effects between routing and deployment features; and (\textbf{\underline{iv}}) a \emph{custom} \texttt{PreferenceWeightedKernel} that wraps $k_c$ to inject structural preferences for model-to-GPU assignments, yielding a preference-weighted deployment kernel $\tilde{k}_c$. Each kernel component is wrapped with a \texttt{ScaleKernel} to learn component-specific output scales, while ARD lengthscales are learned jointly by maximizing the Gaussian process marginal likelihood. The composite kernel ($k_\tau + \tilde{k}_c + k_{\times}$) is integrated into BoTorch~\cite{balandat2020botorch} \texttt{SingleTaskGP} models for system latency and response quality, which are combined via \texttt{ModelListGP}. Outcome transforms are applied as follows: \texttt{Log}$+$\texttt{Standardize} for system latency and \texttt{Logit}$+$\texttt{Standardize} for response quality. All kernels are executed in double precision.

\section{Extended Related Works}
\label{appendix:relatedwork}

\textbf{LLM serving systems.} Recent works have proposed several systems and methods for highly efficient LLM serving~\cite{zhang2025efficient,wang2025thinking,contributors2023lmdeploy,jiang2025cascadia,peng2025hexgen}. AlpaServe~\cite{li2023alpaserve} optimizes system service level objectives (SLOs) by utilizing data and model parallelism; DistServe~\cite{zhong2024distserve} and Splitwise~\cite{patel2024splitwise} split the prefill and decoding phases onto separate GPUs to eliminate interference between them; SarathiServe~\cite{agrawal2024taming} optimizes request batching through prefill chunking to mitigate interference between the prefill and decoding stages; vLLM~\cite{kwon2023efficient} introduces PagedAttention for efficient memory management, enabling higher batch sizes and improved throughput; FastServe~\cite{wu2023fast} employs a preemptive scheduling mechanism that prioritizes shorter jobs to minimize job completion time. These works primarily focus on improving serving performance through techniques such as batching optimization, phase separation, and memory management. In contrast, our work focuses on deployment scheduling with heterogeneous LLMs and GPU types to achieve cost-efficient LLM serving.

\textbf{Heterogeneous model routing.} Model routing systems deploy a router model to dynamically direct incoming queries to different LLMs based on query characteristics, with the primary objective of optimizing system metrics such as cost, latency, or throughput while maintaining output quality. Several recent works have explored this paradigm: RouteLLM~\cite{ong2024routellm} develops cost-optimal routing strategies using preference data and calibrated confidence scores to balance performance and cost; HybridLLM~\cite{ding2024hybrid} proposes an adaptive LLM routing system that switches between large and small models based on query complexity; FrugalGPT~\cite{chenfrugalgpt} cascades LLMs from weakest to strongest, stopping when a model produces a satisfactory response to minimize costs; Router-Bench~\cite{hu2024routerbench} provides a comprehensive benchmark for evaluating routing strategies across diverse query types and model combinations. These works primarily focus on router design and routing strategies for query-level model selection, with limited attention to the routing impact on system performance. In contrast, our work, from a service provider perspective, addresses deployment-level optimization to further enhance system performance.

\textbf{Heterogeneous model deployment.} Heterogeneous model deployment refers to LLM systems that utilize GPUs with varying computational capabilities, memory capacities, or availability characteristics to optimize resource utilization and cost efficiency~\cite{jiang2025demystifying,jiang2025hexgen,yan2024hexiscale,tong2025parallax,yan2025areal}. Among them, HexGen~\cite{jiang2025hexgen} proposes asymmetric model parallelism and evolutionary-based scheduling for distributing computation and communication among heterogeneous GPUs and networks; ThunderServe~\cite{jiang2025thunderserve} accommodates the different computational characteristics of prefill and decoding phases by deploying them on heterogeneous GPUs matched to their respective workload requirements; Helix~\cite{mei2025helix} formulates heterogeneous clusters as a flow network and utilizes a max-flow algorithm to maximize system throughput; SpotServe~\cite{miao2024spotserve} suggests utilizing spot instances for cheap and cost-efficient LLM serving; SkyServe~\cite{mao2025skyserve} provides a multi-cloud serving framework that automatically provisions LLM replicas across different cloud providers and regions to optimize cost and availability. Our work shares a similar motivation: utilizing heterogeneous resources for cost-efficient LLM serving, and we find it naturally complements heterogeneous model routing.

\textbf{Hybrid model parallelism.} Hybrid model parallelism combines tensor parallelism with pipeline parallelism to efficiently scale LLM training and inference beyond what either strategy achieves alone~\cite{zheng2022alpa,li2023alpaserve,miao2022galvatron,jiang2022osdp,wang2024improving,he2025efficient}. Hybrid parallelism enables fitting massive models across multiple GPUs while minimizing inter-machine communication overhead and latency, which is critical for meeting real-time inference requirements.


\end{document}